\documentclass[conference]{IEEEtran}
\usepackage[dvips]{graphicx}
\usepackage[cmex10]{amsmath}
\usepackage{multirow}
\usepackage{epsfig}
\usepackage{mathrsfs}
\usepackage{amssymb}
\usepackage{comment}
\usepackage{bm}
\usepackage{url}
\usepackage{array}
\usepackage{amsthm}
\usepackage{blkarray}
\usepackage{fancyhdr}
\usepackage{enumerate}
\usepackage[lined,boxed,commentsnumbered,ruled,linesnumbered]{algorithm2e}

\renewcommand{\thispagestyle}[1]{} 

\theoremstyle{definition}
\newtheorem{theorem}{Theorem}[section]
\newtheorem{lemma}[theorem]{Lemma}

\ifCLASSINFOpdf

\else

\fi

\usepackage[tight,footnotesize]{subfigure}

\begin{document}
\pagestyle{fancy}
\IEEEoverridecommandlockouts

\lhead{\textit{Technical Report, IBM T. J. Watson Research Center, Yorktown, NY, USA, April, 2015.}}
\rhead{} 
%
\title{Node Failure Localization: Theorem Proof}
\author{\IEEEauthorblockN{Liang Ma\IEEEauthorrefmark{2}, Ting He\IEEEauthorrefmark{2}, Ananthram Swami\IEEEauthorrefmark{4}, Don Towsley\IEEEauthorrefmark{1}, and Kin K. Leung\IEEEauthorrefmark{3}\\}
\IEEEauthorblockA{\IEEEauthorrefmark{2}IBM T. J. Watson Research Center, Yorktown, NY, USA. Email: \{maliang, the\}@us.ibm.com\\
\IEEEauthorrefmark{4}Army Research Laboratory, Adelphi, MD, USA. Email: ananthram.swami.civ@mail.mil\\
\IEEEauthorrefmark{1}University of Massachusetts, Amherst, MA, USA. Email: towsley@cs.umass.edu\\
\IEEEauthorrefmark{3}Imperial College, London, UK. Email: kin.leung@imperial.ac.uk
}
}

\maketitle

\IEEEpeerreviewmaketitle

\section{Introduction}
Selected theorem proof in \cite{Ma15Performance} are presented in detail in this report. We first list the theorems in Section~\ref{sect:theorems} and then give the corresponding proofs in Section~\ref{sect:proofs}. See the original paper \cite{Ma15Performance} for terms and definitions. Table~\ref{t notion} summarizes all graph-theoretical notions used in this report (following the convention in \cite{GraphTheory2005}).

\begin{table}[!htb]
\vspace{-.5em}
\small
\renewcommand{\arraystretch}{1.3}
\caption{Graph-related Notations} \label{t notion}
\vspace{.5em}
\centering
\begin{tabular}{r|m{6.1cm}}
  \hline
  \textbf{Symbol} & \textbf{Meaning} \\
  \hline
  $V$, $L$ & set of nodes/links \\
  \hline
  $M,\: N$ & set of monitors/non-monitors ($M\cup N = V$) \\
  \hline
  $\mathcal{G}-L'$ & delete links: $\mathcal{G}-L'=(V,L\setminus L')$, where ``$\setminus$'' is setminus\\
  \hline
  $\mathcal{G}+L'$ & add links: $\mathcal{G}+L'=(V,L\cup L')$, where the end-points of links in $L'$ must be in $V$\\
  \hline
  $\mathcal{G}+\mathcal{G}'$ & combine two graphs: $\mathcal{G}+\mathcal{G}'=(V(\mathcal{G})\cup V(\mathcal{G}'),L(\mathcal{G})\cup L(\mathcal{G}'))$, where $V(\mathcal{G})$ is the set of nodes and $L(\mathcal{G})$ is the set of links in $\mathcal{G}$\\
  \hline
\end{tabular}
\vspace{-0mm}
\end{table}
\normalsize

\section{Theorems}
\label{sect:theorems}

\begin{lemma}
\label{lemma:RecursiveFunction}
Algorithm~\ref{Alg:MonitorPlacementPolygonlessNetworks} places the minimum number of monitors to ensure the network 1-identifiability in any given connected graph $\mathcal{G}'$, where each biconnected component in $\mathcal{G}'$ (i) has $\beta$ ($\beta=\{0,1,2\}$) neighboring biconnected components, (ii) has $2-\beta$ non-cut-vertex nodes connecting to external monitors (can be outside $\mathcal{G}'$), and (iii) is a PLC.
\end{lemma}

\begin{theorem}
\label{theorem:OMP-CSP}
OMP-CSP ensures that any single-node failure in a given network is uniquely identifiable under CSP using the minimum number of monitors.
\end{theorem}

\section{Proofs}
\label{sect:proofs}

\subsection{Proof of Lemma~\ref{lemma:RecursiveFunction}}
\label{proof:lemma:RecursiveFunction}

\addtocounter{algocf}{+2}
\begin{algorithm}[tb]
\small
\SetKwInOut{Input}{input}\SetKwInOut{Output}{output}
\Input{Network topology $\mathcal{G}$, node set $S$}
\Output{Sub-set of nodes in $\mathcal{G}$ as monitors}
\If{$|L|= 0$}
{return\;}
\ForEach{connected component $\mathcal{G}_i$ in $\mathcal{G}$}
{
\uIf{$\mathcal{G}_i$ contains only one biconnected component}
    {randomly choose a node in $\mathcal{G}_i$ as a monitor\;\label{RandOneBC}}
\Else
    {in $\mathcal{G}_i$, label one biconnected component with 0 or 1 cut-vertex as $\mathcal{B}_1$, one neighboring biconnected component of $\mathcal{B}_1$ as $\mathcal{B}_2$ (if any), and one neighboring biconnected component of $\mathcal{B}_2$ other than $\mathcal{B}_1$ as $\mathcal{B}_3$ (if any)\;\label{assignNumber}
   \uIf{$\mathcal{B}_2$ is a bond}
     {choose the common node between $\mathcal{B}_1$ and $\mathcal{B}_2$ as a monitor\;\label{deployCase2}
     $\mathcal{G}'_i\leftarrow \mathcal{G}_i\ominus(\mathcal{B}_1+\mathcal{B}_2)$\;\label{removeCase2}}
   \Else(\texttt{\small //$\mathcal{B}_2$ is not a bond})
     {randomly choose node $v$ ($v\notin S$) in $\mathcal{B}_2$ as a monitor\;\label{deployCase3}
     $\mathcal{G}'_i\leftarrow \mathcal{G}_i\ominus(\mathcal{B}_1+\mathcal{B}_2+\mathcal{B}_3)$ (if $\mathcal{B}_3$ exists)\;\label{removeCase3}}
     \emph{Monitors-in-Polygon-less-Network}($\mathcal{G}'_i$, $S$)\;\label{recursiveCall}}
}
\caption{Monitors-in-Polygon-less-Network($\mathcal{G}$, $S$)}
\label{Alg:MonitorPlacementPolygonlessNetworks}
\vspace{-.1em}
\end{algorithm}
\normalsize

\begin{figure}[tb]
\centering
\includegraphics[width=.99\columnwidth]{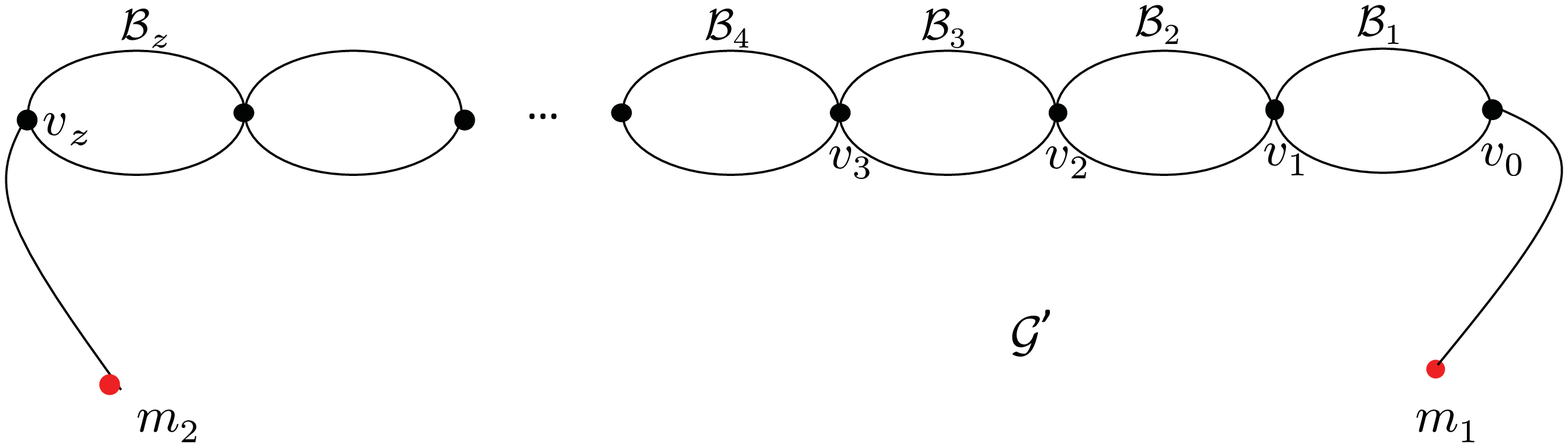}
\caption{Necessary monitor placement in tandem networks.}\label{fig:MonitorLinearNetwork}
\end{figure}

For Algorithm~\ref{Alg:MonitorPlacementPolygonlessNetworks}, the input network is not necessarily a connected graph. Lemma~\ref{lemma:RecursiveFunction}, however, only considers the case that the input network satisfying the three conditions (in Lemma~\ref{lemma:RecursiveFunction}) is connected. Then it suffices to show that Algorithm~\ref{Alg:MonitorPlacementPolygonlessNetworks} places the minimum number of monitors to ensure that any two non-monitors in $\mathcal{G}'$ (satisfying the three conditions in Lemma~\ref{lemma:RecursiveFunction}) are distinguishable. Such connected input network can be represented as a tandem network\footnote{In this report, networks with such structures are called \emph{tandem networks}.}, shown in Fig.~\ref{fig:MonitorLinearNetwork}. As illustrated in Fig.~\ref{fig:MonitorLinearNetwork}, suppose there are $z$ biconnected components in $\mathcal{G}'$, where $\mathcal{B}_1$ and $\mathcal{B}_z$ have external monitor connections via $v_0$ and $v_z$ to $m_1$ and $m_2$ outside $\mathcal{G}'$.

(1) We first prove that any non-cut-vertex (excluding the nodes connecting to external monitors, e.g., $v_0$ and $v_z$ in Fig.~\ref{fig:MonitorLinearNetwork}), denoted by $w$, in $\mathcal{G}'$ is 1-identifiable if $\exists$ another monitor, denoted by $m_3$, in $\mathcal{G}'$ ($m_3$ can be anywhere in $\mathcal{G}'$).

\emph{(1.a) The case that $m_1\neq m_2$.}
Suppose $w$ is in biconnected component $\mathcal{B}_i$. Then according to Theorem~15 \cite{Ma15Performance}, $w$ is distinguishable from any other node (including cut-vertices or nodes connecting to external monitors) in $\mathcal{B}_i$. Moreover, for a node outside $\mathcal{B}_i$, say $u$, it is impossible that measurement path (from $m_1$ to $m_2$ in Fig.~\ref{fig:MonitorLinearNetwork}) must go through $u$ and $w$ at the same time, since $w$ must be a cut-vertex in Fig.~\ref{fig:MonitorLinearNetwork} otherwise, contradicting the assumption. Therefore, $w$ is also distinguishable from nodes outside $\mathcal{B}_i$. Thus, $w$ is 1-identifiable when $m_1\neq m_2$ even without $m_3$.

\emph{(1.b) The case that $m_1= m_2$.}
To ensure each node in $\mathcal{G}'$ is 1-identifiable, at least one extra monitor (besides $m_1$ and $m_2$) that can generate simple measurement paths traversing $\mathcal{G}'$ is required. Now we have monitor $m_3$ in $\mathcal{G}'$. In this case, $\mathcal{G}'$ can be further decomposed into subgraphs. Then the argument in \emph{(1.a)} applies to each subgraph by using $m_3$.

In sum, $w$ is 1-identifiable in $\mathcal{G}'$, i.e., any non-cut-vertex (excluding the nodes connecting to external monitors) in $\mathcal{G}'$ is 1-identifiable if $\exists$ a monitor in $\mathcal{G}'$.

(2) Based on the argument in (1), we know that additional monitor placement (besides $m_1$ and $m_2$ in Fig.~\ref{fig:MonitorLinearNetwork}) in $\mathcal{G}'$ is only for distinguishing the cut-vertices and nodes connecting to external monitors. Now we consider how to place the minimum number (at least one monitor, thus the argument in \emph{(1.b)} still holds) of monitors to distinguish all these nodes in $\mathcal{G}'$. In Algorithm~\ref{Alg:MonitorPlacementPolygonlessNetworks}, line~\ref{assignNumber} assigns a sequence number to each biconnected component. This is feasible as $\mathcal{G}'$ is a tandem network. Suppose $|\mathcal{B}_i|\geq 3$ (i.e., $\mathcal{B}_i$ is not bond). Then, as Fig.~\ref{fig:MonitorLinearNetwork} illustrates, to ensure the 1-identifiability of $v_0$, there are three possible locations for necessary monitor placement: (i) $v_0$ connects to another external monitor; (ii) place a monitor in $\mathcal{B}_1$; or (iii) select a monitor from $V(\mathcal{B}_2)\setminus \{v_2\}$. Similarly, there are also three possible locations for monitor placement such that $v_1$ and $v_2$ are 1-identifiable. To ensure the 1-identifiability of $v_0$, $v_1$, and $v_2$, we notice that there exists a common location, i.e., a node in $V(\mathcal{B}_2)\setminus \{v_1 , v_2\}$, where placing one monitor can guarantee that $v_0$, $v_1$, and $v_2$ are all 1-identifiable. Moreover, no other places can guarantee that $v_0$, $v_1$, and $v_2$ are 1-identifiable at the same time. Thus, line~\ref{deployCase3} selects a monitor, denoted $m_3$, from $V(\mathcal{B}_2)\setminus \{v_1 , v_2\}$. Note that the selection of $m_3$ only guarantees that $v_0$, $v_1$, and $v_2$ are 1-identifiable, i.e., the identifiability of all other nodes remain the same. Hence, $m_3$ does not affect the necessity of previously deployed monitors. However, if $\mathcal{B}_2$ is a bond and $\mathcal{B}_1$ is not a bond, then to ensure that $v_0$ is 1-identifiable, there are only two possible locations for monitor placement: (i) $v_0$ connects to another external monitor or (ii) place a monitor in $\mathcal{B}_1$. Meanwhile, to ensure that $v_1$ is 1-identifiable, the possible monitor locations are also reduced to two: (i) place a monitor in $\mathcal{B}_1$ or (ii) place a monitor in $\mathcal{B}_3$. In such case, the common location to ensure the 1-identifiability of both $v_0$ and $v_1$ is in $\mathcal{B}_1$. Thus, line~\ref{deployCase2} deploys a monitor at the common node between $\mathcal{B}_1$ and $\mathcal{B}_2$. As aforementioned, for this newly selected monitor, it is only for identifying a specific set of nodes. All previously deployed monitors remain necessary. After this placement, lines~\ref{removeCase2} and \ref{removeCase3} remove the processed biconnected components. Then the remaining graph is processed by Algorithm~\ref{Alg:MonitorPlacementPolygonlessNetworks} recursively, as further monitor placement is independent of the monitors that are already deployed. Finally, one trivial case we have not discussed is that $\mathcal{G}_i$ contains only one biconnected component, where a randomly chosen monitor (line~\ref{RandOneBC}) can ensure the network 1-identifiability. Therefore, for the input network satisfying the conditions in Lemma~\ref{lemma:RecursiveFunction}, Algorithm~\ref{Alg:MonitorPlacementPolygonlessNetworks} can place the minimum number of monitors for achieving the network 1-identifiability.
\hfill$\blacksquare$

\subsection{Proof of Theorem~\ref{theorem:OMP-CSP}}

\addtocounter{algocf}{-2}
\begin{algorithm}[tb]
\small
\SetKwInOut{Input}{input}\SetKwInOut{Output}{output}
\Input{Connected network topology $\mathcal{G}$ which is not 2-connected}
\Output{Set of monitors that achieves the 1-identifiability in $\mathcal{G}$ under CSP}
partition $\mathcal{G}$ into biconnected components $\{\mathcal{B}_1, \mathcal{B}_2, \ldots\}$ and then PLCs\;
\If{$\mathcal{G}$ is 2-connected}
    {
    deploy monitors by \emph{Monitors-in-Biconneted-Network} (Algorithm~\ref{Alg:MonitorPlacement-CSP-2connectivity} in \cite{Ma15Performance})\;\label{OMPauxiliary}
    return\;
    }
\ForEach{biconnected component $\mathcal{B}_i$\label{ProcessAllBC}}
{
\uIf{$\mathcal{B}_i$ is a PLC \textbf{and} $\mathcal{B}_i$ has only one cut-vertex}
   {
   randomly choose node $v$ ($v$ is not a cut-vertex in $\mathcal{G}$) in $\mathcal{B}_i$ as a monitor\;\label{OMPnecessaryBCPLC}
   }
\Else
   {
   find set $A$ containing all PLCs with $\geq 3$ agents or $\geq 4$ neighboring PLCs within $\mathcal{B}_i$, and set $C$ containing all neighboring PLCs of each PLC in set $A$ within $\mathcal{B}_i$\;\label{OMPidentifiableNodesBC3}
   within $\mathcal{B}_i$, find set $E$ containing all PLCs with only 2 agents and one agent is a cut-vertex (in $\mathcal{G}$)\;\label{OMPidentifiableNodesBC4}
   $\mathcal{B}'_i\leftarrow \mathcal{B}_i\ominus (A\cup C\cup E)$\;\label{OMPremainingInBC}
   \emph{Monitors-in-Polygon-less-Network}($\mathcal{B}'_i$, $S^i_a$), where $S^i_a$ is the set of agents in $\mathcal{B}_i$\;\label{OMPnecessaryBCend}
   }
}\label{ProcessAllBCend}
find set $F$ containing all biconnected components with monitors\;\label{OMPidentifiableNodes1}
find set $I$ containing all biconnected components with 2 cut-vertices (in $\mathcal{G}$) and 3 or more neighboring biconnected components in $\mathcal{G}$\;\label{OMPidentifiableNodes3}
find set $J$ containing all biconnected components with 3 or more cut-vertices, and set $K$ containing all neighboring biconnected components of each component in $J$\;\label{OMPidentifiableNodes5}
$\mathcal{G}'\leftarrow \mathcal{G}\ominus (F\cup I \cup J \cup K)$\;\label{OMPfinalRemaining}
\emph{Monitors-in-Polygon-less-Network}($\mathcal{G}'$, $S_c$), where $S_c$ is the set of cut-vertices in $\mathcal{G}$\;\label{OMPfinalPlacement}
\caption{Optimal Monitor Placement for 1-identifiability under CSP (OMP-CSP)}
\label{Alg:MonitorPlacement-CSP-1connectivity}
\vspace{-.1em}
\end{algorithm}
\normalsize

First, we consider the case that the input connected network is not 2-connected. In this case, there exist at least one cut-vertex and two biconnected components, and auxiliary algorithm Algorithm~\ref{Alg:MonitorPlacement-CSP-2connectivity} is not invoked. For such input network, we discuss it as follows.

(1) We first place necessary monitors in each biconnected component. If biconnected component $\mathcal{B}_i$ has only one cut-vertex, denoted by $v_c$, then at least one non-cut-vertex in $\mathcal{B}_i$ should be a monitor.

\emph{(1.a)} If $\mathcal{B}_i$ is a PLC, then we can randomly select a non-cut-vertex, denoted by $m_i$, as a monitor (line~\ref{OMPnecessaryBCPLC}) according to Theorem~15 and Corollary~16 \cite{Ma15Performance}, such that any node $w$ with $w\in V(\mathcal{B}_i)\setminus \{v_c\}$ is distinguishable from any node in $V(\mathcal{B}_i) $. Moreover, $w$ is also distinguishable from nodes outside $\mathcal{B}_i$ due to the existence of path from $m_i$ to $v_c$ without traversing $w$. Therefore, all nodes in $V(\mathcal{B}_i)\setminus \{v_c\}$ are 1-identifiable when $\mathcal{B}_i$ is a PLC and has only one cut-vertex.

\emph{(1.b)} If $\mathcal{B}_i$ is not a PLC, i.e., $\mathcal{B}_i$ contains at least one polygon, then the required monitor in $\mathcal{B}_i$ cannot be randomly placed. For this case, we need to first find all nodes in $\mathcal{B}_i$ that are guaranteed to be 1-identifiable. For these 1-identifiable nodes, there are three cases:

\begin{figure}[tb]
\centering
\includegraphics[width=.6\columnwidth]{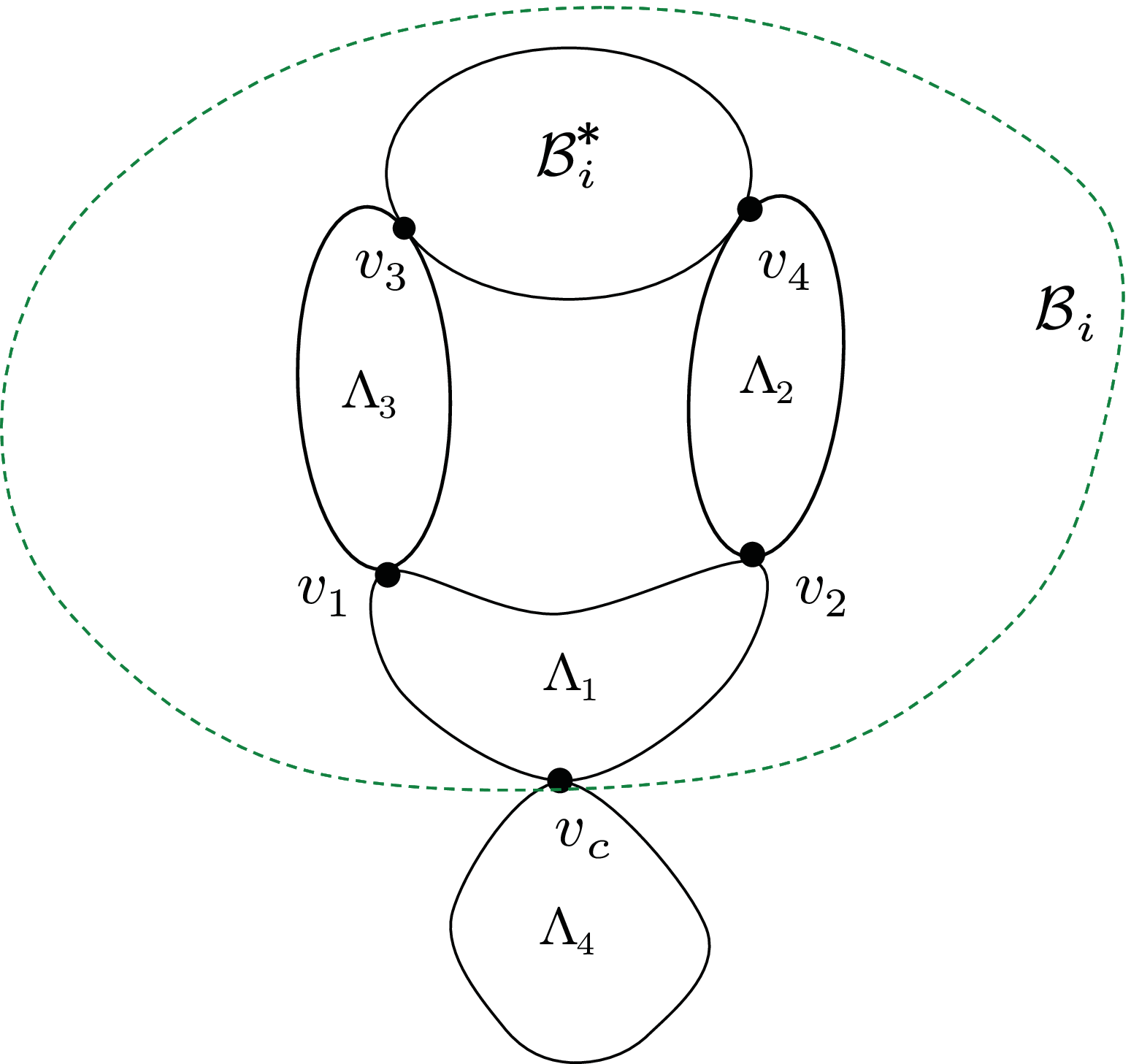}
\caption{PLC $\Lambda_1$ with three agents.}\label{fig:OneVC3Agents}
\end{figure}
\begin{figure}[tb]
\centering
\includegraphics[width=.95\columnwidth]{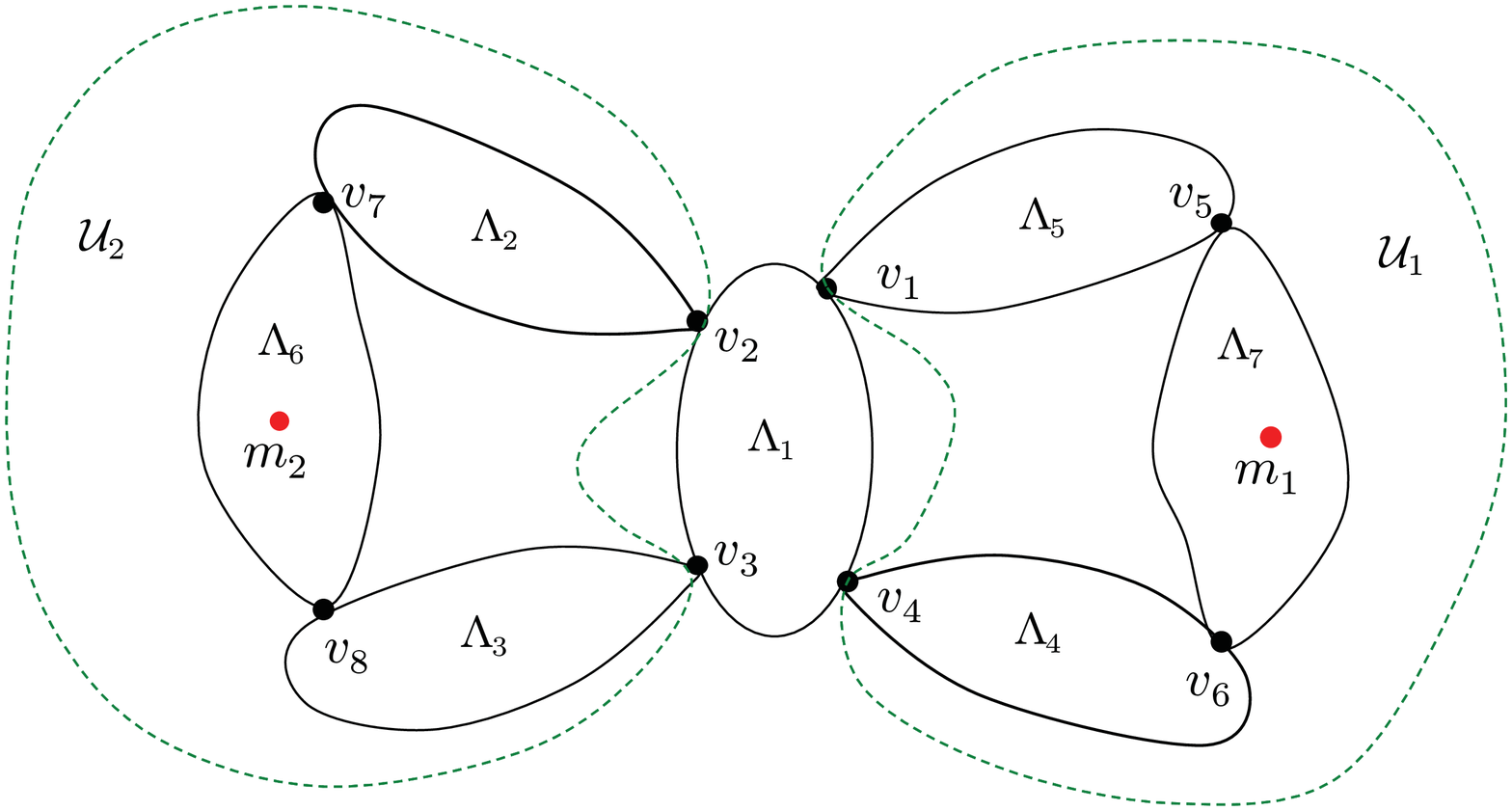}
\caption{PLC $\Lambda_1$ with four neighboring PLCs.}\label{fig:PLC4Neighbors}
\end{figure}
\begin{figure}[tb]
\centering
\includegraphics[width=.7\columnwidth]{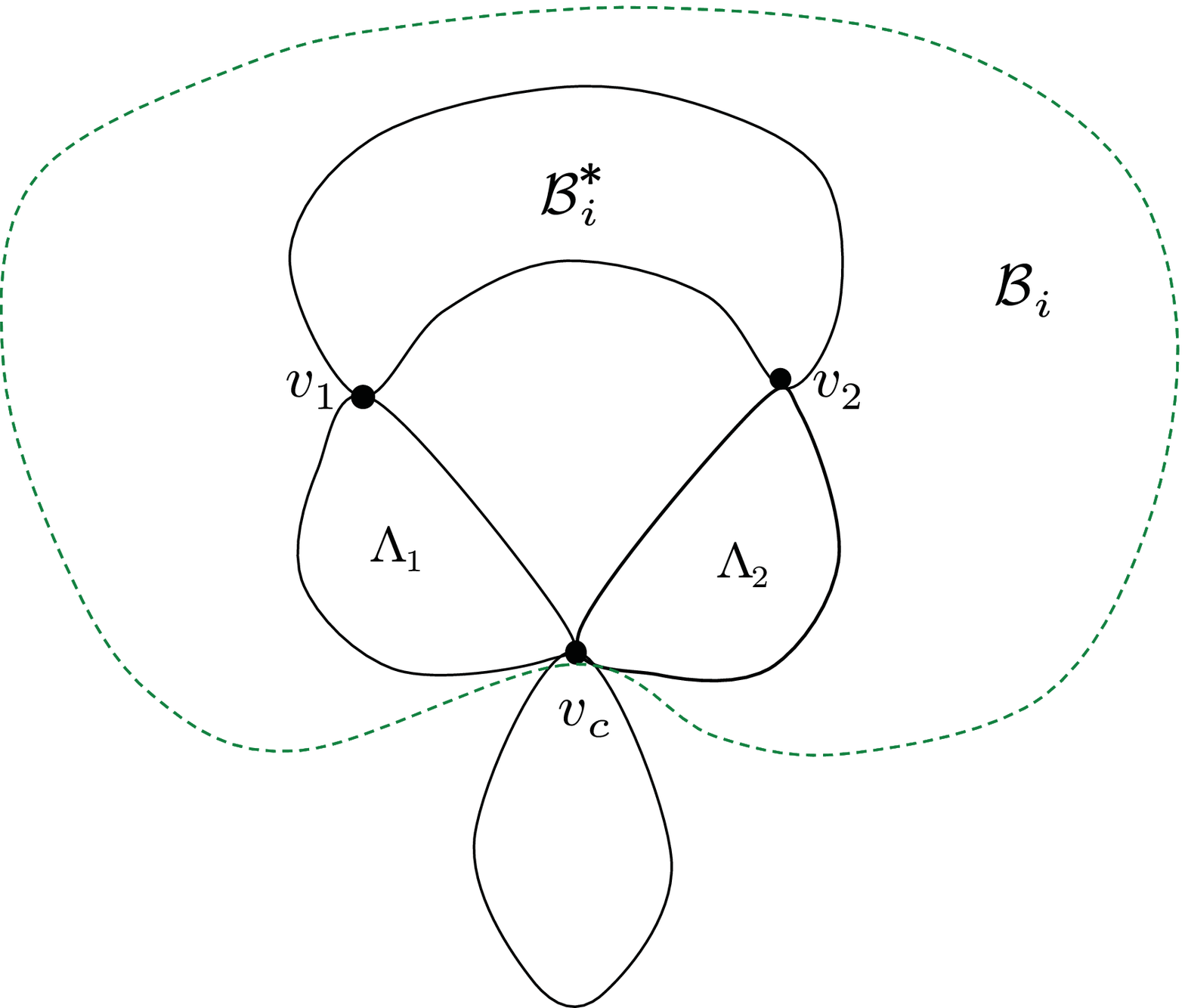}
\caption{PLC $\Lambda_1$ (and $\Lambda_2$) with two agents and one agent is a cut-vertex.}\label{fig:OneVC2Agents}
\end{figure}

\begin{enumerate}
  \item[\textcircled{\small 1}]\normalsize A PLC with 3 or more agents. As shown in Fig.~\ref{fig:OneVC3Agents}, PLC $\Lambda_1$ has three agents and $\mathcal{B}_i$ must have a monitor (say $m_{\mathcal{B}_i}$) that is not $v_c$. Moreover, outside $\mathcal{B}_i$, $v_c$ must connect to a monitor, e.g., a monitor in $\Lambda_4$ (denoted by $m_{\Lambda_4}$). Within $\Lambda_1$, for any two nodes $w_1$ and $w_2$ ($w_1\neq w_2 \neq v_c$), we can find a path traversing only $w_1$ but not $w_2$ (and vice versa) using $m_{\mathcal{B}_i}$ and $m_{\Lambda_4}$ as $\Lambda_1$ is a PLC. Therefore, each node in $V(\Lambda_1)\setminus \{v_c\}$ is 1-identifiable. Then following the similar argument in (1) of Section~\ref{proof:lemma:RecursiveFunction}, we know each node in $V(\Lambda_3+\Lambda_2)\setminus \{v_1,v_2,v_3,v_4\}$ is 1-identifiable, where $\Lambda_2$ and $\Lambda_3$ are the neighboring PLCs of $\Lambda_1$. Therefore, to ensure that all non-cut-vertices are 1-identifiable, monitor placement in $\mathcal{B}_i$ only needs to make sure that nodes in $\mathcal{B}^*_i$ (see Fig.~\ref{fig:OneVC3Agents}) are 1-identifiable. However, $\mathcal{B}^*_i$ may not be a tandem network, i.e., $\mathcal{B}^*_i$ possibly contains a polygon, which needs to be further processed in the following cases.
  \item[\textcircled{\small 2}]\normalsize A PLC with 4 or more neighboring PLCs. This case is illustrated in Fig.~\ref{fig:PLC4Neighbors}. In Fig.~\ref{fig:PLC4Neighbors}, $\Lambda_1$ has two neighboring polygons, which implies that $\Lambda_1$ has 4 or more neighboring PLCs. In this case, $\mathcal{U}_1$ must have at least one monitor, since $v_5$ and $v_6$ are not distinguishable otherwise. Similarly, $\mathcal{U}_2$ must also have a monitor. Using these two monitors, every node in $\Lambda_1$ can be shown to be 1-identifiable as $\Lambda_1$ is a PLC. Then following the similar argument in (1) of Section~\ref{proof:lemma:RecursiveFunction}, we know each node in $V(\Lambda_2+\Lambda_3+\Lambda_4+\Lambda_5)\setminus \{v_j\}^{8}_{j=5}$ is 1-identifiable, where $\Lambda_2$, $\Lambda_3$, $\Lambda_4$, and $\Lambda_5$ are the neighboring PLCs of $\Lambda_1$.
  \item[\textcircled{\small 3}]\normalsize A PLC with 2 agents and one agent is a cut-vertex. This case is shown in Fig.~\ref{fig:OneVC2Agents}. Following the similar argument in (1) of Section~\ref{proof:lemma:RecursiveFunction}, we know that each node in $V(\Lambda_1+\Lambda_2)\setminus \{v_1,v_2,v_c\}$ is 1-identifiable, where $\Lambda_1$ and $\Lambda_2$ have the common node $v_c$. However, $v_1$ and $v_2$ may or may not be distinguishable depending on the topology of $\mathcal{B}^*_i$.
\end{enumerate}
Lines~\ref{OMPidentifiableNodesBC3}--\ref{OMPidentifiableNodesBC4} consider the above three cases to remove all 1-identifiable nodes\footnote{$v_c$ in Fig.~\ref{fig:OneVC3Agents} and Fig.~\ref{fig:OneVC2Agents} is also temporarily removed; however, it still exists in neighboring biconnected components, and thus the 1-identifiability of $v_c$ will be considered later in lines~\ref{OMPidentifiableNodes1}--\ref{OMPfinalPlacement}.}. Note that in lines~\ref{OMPidentifiableNodesBC3}--\ref{OMPidentifiableNodesBC4}, we get sets $A$, $C$, and $E$, the common non-cut-vertices among these three sets may not be marked as 1-identifiable in any of the above three cases. Nevertheless, we can prove these common non-cut-vertices are also 1-identifiable as follows: Let $U_c$ be the set containing all such common non-cut-vertices, and $U_r=A \cup C \cup E\setminus U_c$. Now consider a random node $z$ with $z\in U_c$ and another node $x$. There are 3 cases: (i) If $x\in U_r$, then we know $x$ is distinguishable from $z$ based on previous results; (ii) if $x\in U_c$, then $\exists$ a path traversing $x$ without going through $z$ using nodes in sets $A$, $C$, and $E$; (iii) if $x\notin U_r\cup U_c$, then as (ii) shows that $z$ and $x$ are also distinguishable because of the existence of paths bypassing $z$ using nodes in $U_r\cup U_c$. Therefore, the common non-cut-vertex $z$ is 1-identifiable. Using this union set $A\cup C\cup E$, the remaining graph obtained by line~\ref{OMPremainingInBC} is a collection of tandem networks. Within this collection of tandem networks, consider two non-monitors $w_1$ and $w_2$ in two different connected tandem networks. We can show that $w_1$ and $w_2$ are distinguishable. This is because each connected tandem network must have additional necessary monitors. Using these additional monitors and also the removed components, we can find paths traversing only $w_1$ or $w_2$. Moreover, for these additional monitors, each has at least two internally vertex disjoint paths to any cut-vertex in the parent biconnected component. Thus, each non-monitor in one of these connected tandem network is distinguishable from any non-monitor outside its parent biconnected component. Therefore, it suffices to only consider how to enable the 1-identifiability in each of these connected tandem networks. This goal can be achieved by Algorithm~\ref{Alg:MonitorPlacementPolygonlessNetworks}, the correctness of which is shown in Lemma~\ref{lemma:RecursiveFunction}.

(2) In processing each biconnected component, we only place the necessary monitors. These necessary monitor placements are proved to be able to ensure that all non-cut-vertex nodes in biconnected components are 1-identifiable. Next, we can consider the 1-identifiability of cut-vertices. Note that for a biconnected component, if no necessary monitors are placed so far, that means this biconnected component is either a PLC or contains a sufficient number of cut-vertices so that no additional monitors are required. For these biconnected components without monitors, it is still possible to find 1-identifiable cut-vertices in the following two cases:

\begin{figure}[tb]
\centering
\vspace{-.5em}
\includegraphics[width=.55\columnwidth]{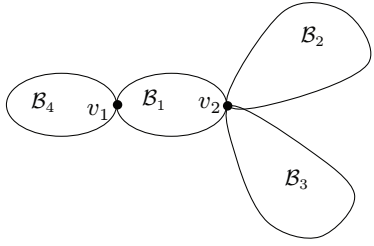}
\caption{Biconnected component $\mathcal{B}_1$ with two cut-vertices and three neighboring biconnected components.}\label{fig:TwoCV3BC}
\vspace{-.5em}
\end{figure}
\begin{figure}[tb]
\centering
\vspace{-.5em}
\includegraphics[width=.7\columnwidth]{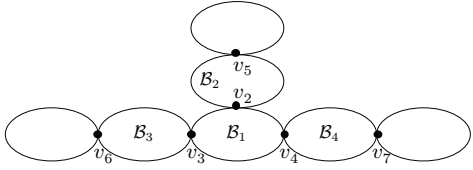}
\caption{Biconnected component $\mathcal{B}_1$ with three cut-vertices and three neighboring biconnected components.}\label{fig:ThreeCV3BC}
\vspace{-.5em}
\end{figure}

\begin{enumerate}
  \item[\textcircled{\small 1}]\normalsize A biconnected component with 2 cut-vertices and 3 or more neighboring biconnected components, such as $\mathcal{B}_1$ in Fig.~\ref{fig:TwoCV3BC}. In this case, we know that $v_2$ must have two monitor connections, one through $\mathcal{B}_2$ and the other through $\mathcal{B}_3$. Meanwhile, $v_1$ also has a monitor connection through $\mathcal{B}_4$. Using these monitor connections, each node in $V(\mathcal{B}_1)\setminus \{v_1\}$ is 1-identifiable. Note that unlike our previous discussion in (1) that connecting point may not be 1-identifiable. In Fig.~\ref{fig:TwoCV3BC}, cut-vertex $v_2$ is guaranteed to be 1-identifiable, because it has more than two internally vertex disjoint monitor connections.
  \item[\textcircled{\small 2}]\normalsize A biconnected component with 3 or more cut-vertices, such as $\mathcal{B}_1$ in Fig.~\ref{fig:ThreeCV3BC}. In this case, $v_i$ ($i=\{2,3,4\}$) must have a monitor connection through $\mathcal{B}_i$. Using these monitor connections, all nodes in $\mathcal{B}_1$ (including the cut-vertices) are 1-identifiable. Similarly, nodes in $V(\mathcal{B}_2+\mathcal{B}_3+\mathcal{B}_4)\setminus \{v_5,v_6,v_7\}$ are also 1-identifiable, where $\mathcal{B}_2$, $\mathcal{B}_3$, and $\mathcal{B}_4$ are neighboring biconnected components of $\mathcal{B}_1$.
\end{enumerate}
Lines~\ref{OMPidentifiableNodes3}--\ref{OMPidentifiableNodes5} consider all above cases to determine the 1-identifiable nodes. Removing the 1-identifiable nodes by line~\ref{OMPfinalRemaining}, we get a collection of tandem networks without containing any monitors. Following our previous arguments about monitor placement in a collection of tandem networks (arguments after \textcircled{\small 3} \normalsize  in (1)), we further deploy monitors optimally by lines~\ref{OMPfinalRemaining}--\ref{OMPfinalPlacement}.

In this way, we use the minimum number of monitors to ensure that any cut-vertex is 1-identifiable in a given non-2-connected network.

\newcounter{hdps}
\stepcounter{hdps}

\begin{algorithm}[tb]
\renewcommand\thealgocf{\Alph{hdps}}
\small
\SetKwInOut{Input}{input}\SetKwInOut{Output}{output}
\Input{2-connected network $\mathcal{G}$ and its PLCs}
\Output{Set of monitors that achieves 1-identifiability in $\mathcal{G}$ under CSP}
\If{$\mathcal{G}$ is a PLC}
    {randomly select two nodes as monitors;
    return\;\label{AuxEasiestCase}}
\uIf{$\exists$ non-empty set $A$ containing all PLCs with 4 or more neighboring PLCs within $\mathcal{G}$\label{AuxObser3-1}}
    {find set $C$ containing all neighboring PLCs of each PLC in set $A$ within $\mathcal{G}$\;\label{AuxObser3-2}
    $\mathcal{G}'\leftarrow \mathcal{G}\ominus (A\cup C)$\;\label{AuxRemainingCase1}
    \emph{Monitors-in-Polygon-less-Network}($\mathcal{G}'$, $S_a$), where $S_a$ denotes the set of agents in $\mathcal{G}$\;\label{AuxfinalPlacement1}
    }\texttt{\small //In the following cases, $\mathcal{G}$ must contain only one polygon}\;
\uElseIf{all PLCs in $\mathcal{G}$ are non-bonds\label{senario1}}
    {
    randomly select a non-agent node in a PLC, denoted by $\Lambda$, as a monitor\;\label{AuxNoBonds}
    $\mathcal{G}'\leftarrow \mathcal{G}\ominus (\{\Lambda\}\cup E)$, where $E$ is the set containing all neighboring PLCs of $\Lambda$\;\label{AuxRemainingCase2}
    \emph{Monitors-in-Polygon-less-Network}($\mathcal{G}'$, $S_a$)\;\label{AuxfinalPlacement2}
    }
\Else(\texttt{\small //$\exists$ at least one bond PLC}\label{senario2})
    {
    randomly select a bond PLC $\Lambda'$ with two end-points $v_1$ and $v_2$ and two neighboring PLCs $\Lambda_1$ ($v_1\in\Lambda_1$) and $\Lambda_2$ ($v_2\in\Lambda_2$)\;\label{AuxSelectBond}
    \uIf{$\Lambda_1$ ($\Lambda_2$) is a bond\label{AuxSelectCandidate}}
    {$w_1\leftarrow v_1$ ($w_2\leftarrow v_2$)\;}
    \Else
    {$w_1$ ($w_2$) $\leftarrow$ a (random) non-agent node in $\Lambda_1$ ($\Lambda_2$)\;}\label{AuxSelectCandidateEnd}
    \ForEach{$i=1,2$\label{AuxTest}}
    {
    select $w_i$ as a monitor\;
    $\mathcal{G}'_i \leftarrow \mathcal{G}\ominus \Gamma_{w_i}$, where $\Gamma_{w_i}$ is the set involving (i) PLCs that contain $w_i$, and (ii) neighboring PLCs of the PLCs in (i) if $w_i$ is not an agent\;
    \emph{Monitors-in-Polygon-less-Network}($\mathcal{G}'_i$, $S_a$)\;
    }\label{AuxTestEnd}
    in above two monitor placements, select the one with the minimum number of monitors as the final output\;\label{AuxSelectTwoTests}
    }\label{AuxLastStep}
\caption{\emph{Monitors-in-Biconneted-Network}}
\label{Alg:MonitorPlacement-CSP-2connectivity}
\vspace{-.1em}
\end{algorithm}
\normalsize

Finally, we consider the case that the input connected network is 2-connected. In this case, there are no cut-vertices. However, it is still possible to determine some 1-identifiable nodes. Specifically, Case-\textcircled{\small 2} \normalsize in (1) can be applied to 2-connected network $\mathcal{G}$ with 2 or more polygons. This particular case is handled by line~\ref{AuxObser3-1}--\ref{AuxfinalPlacement1} in Algorithm~\ref{Alg:MonitorPlacement-CSP-2connectivity}. However, if no 1-identifiable nodes can be found, then it implies that 2-connected $\mathcal{G}$ itself is a PLC or contains one and only one polygon. If the given 2-connect network is a PLC, then randomly selecting two monitors (line~\ref{AuxEasiestCase} of Algorithm~\ref{Alg:MonitorPlacement-CSP-2connectivity}) can ensure network 1-identifiability according to Theorem~15 \cite{Ma15Performance}. While for a 2-conencted network with one and only one polygon, our strategy is to deploy the first monitor, remove the 1-identifiable nodes using our previous methods in Algorithm~\ref{Alg:MonitorPlacement-CSP-1connectivity}, and then apply Algorithm~\ref{Alg:MonitorPlacementPolygonlessNetworks} to optimally deploy monitors in the remaining graph. For a 2-connected network with only one polygon, there are two cases:

\begin{figure}[tb]
\centering
\vspace{-.5em}
\includegraphics[width=.35\columnwidth]{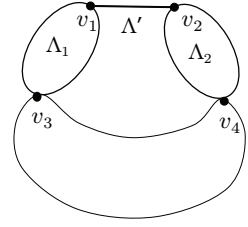}
\caption{Monitor placement in 2-connected networks.}\label{fig:BCbond}
\vspace{-.5em}
\end{figure}

\begin{enumerate}
  \item[\textcircled{\small 1}]\normalsize All PLCs in $\mathcal{G}$ are non-bonds. For this case, there is no difference in selecting which PLC to deploy the first monitor because all PLCs have the same structure with at least one non-agent node. This case is captured by lines~\ref{senario1}--\ref{AuxfinalPlacement2} in Algorithm~\ref{Alg:MonitorPlacement-CSP-2connectivity}.
  \item[\textcircled{\small 2}]\normalsize $\exists$ at least one bond in $\mathcal{G}$. In this case, we randomly select a bond PLC, denoted by $\Lambda'$ with two end-points $v_1$ and $v_2$ and two neighboring PLCs $\Lambda_1$ ($v_1\in\Lambda_1$) and $\Lambda_2$ ($v_2\in\Lambda_2$), shown in Fig.~\ref{fig:BCbond}. To distinguish $v_1$ and $v_2$ in Fig.~\ref{fig:BCbond}, we need a monitor in $\Lambda_1$ or $\Lambda_2$ or both. Depending on if $\Lambda_1$ or $\Lambda_2$ is a bond, we have to select $v_1$ or $v_2$ as a monitor. To get such monitor candidates, we use lines~\ref{AuxSelectCandidate}--\ref{AuxSelectCandidateEnd} to get $w_1$ and $w_2$ for possibly placing the first monitor. Unfortunately, we have no knowledge on which one (selecting $w_1$ or $w_2$ as a monitor) can generate the optimal solution. Therefore, we test them both, and select the one with the minimum number of monitors as the final output; see lines~\ref{AuxTest}--\ref{AuxSelectTwoTests} of Algorithm~\ref{Alg:MonitorPlacement-CSP-2connectivity}.
\end{enumerate}
In all, the above discussion on 2-connected input network is complete to cover all cases of 2-connected networks.

Consequently, OMP-CSP (Algorithm~\ref{Alg:MonitorPlacement-CSP-1connectivity}) can guarantee network 1-identifiability using the minimum number of monitors for any given network topology.
\hfill$\blacksquare$

\bibliographystyle{IEEEtran}
\bibliography{mybibSimplifiedA}
\end{document}